\documentclass[twoside]{dis08}
\usepackage[latin1]{inputenc}
\usepackage[dvips]{graphicx}
\usepackage{amssymb,amsmath}

\begin{document}
\title{{\boldmath Gluon-induced QCD Corrections to $pp \to ZZ \to \ell\bar{\ell}\ell'\bar{\ell'}$}}

\author{T.~Binoth$^1$, \underline{N.~Kauer}$^2$ and P.~Mertsch$^3$
\vspace{.3cm}\\
$^1$School of Physics, The University of Edinburgh,\\
Edinburgh EH9 3JZ, United Kingdom
\vspace{.1cm}\\
$^2$Institut f\"ur Theoretische Physik, Universit\"at W\"urzburg,\\
D-97074 W\"urzburg, Germany
\vspace{.1cm}\\
$^3$Rudolf Peierls Centre for Theoretical Physics, University of Oxford,\\
Oxford OX1 3NP, United Kingdom\\
}

\maketitle

\begin{abstract}
A calculation of the loop-induced gluon-fusion process 
$gg \to Z^\ast(\gamma^\ast)Z^\ast(\gamma^\ast) \to \ell\bar{\ell}\ell'\bar{\ell'}$ 
is presented, 
which provides an important background for Higgs boson searches in the $H \to ZZ$
channel at the LHC. 
We find that the photon contribution is important for Higgs masses below the 
$Z$-pair threshold and that the $gg$-induced process yields a correction of about 
$15\%$ relative to the NLO QCD prediction for the $q\bar{q}$-induced process when 
only a $M_{\ell\bar{\ell}}, M_{\ell'\bar{\ell'}} > 5$ GeV cut is applied.
\end{abstract}


\section{Introduction}

Accurate theoretical predictions for the hadronic production of vector boson pairs 
are needed not only for tests of the non-Abelian gauge structure
of the Standard Model, but also to determine an important background to 
Higgs boson searches at the LHC
\cite{kauer_nikolas:leshouches_higgs,Dawson:2008uu,Kauer:2007zz}.
Due to the large gluon flux at the LHC the contribution from gluon-gluon and 
gluon-quark scattering is enhanced.  In vector boson pair
production such subprocesses do not contribute at leading order (LO).
In LHC Higgs searches higher order corrections to background predictions can be 
further enhanced by experimental selection cuts.  For example, the $gg$-induced 
subprocess to $pp \to WW\to\ell\bar{\nu}\bar{\ell'}\nu'$, which 
contributes formally at next-to-next-to-leading order QCD, gives a 
30\% correction to the next-to-leading order (NLO) QCD prediction when realistic 
Higgs search selection cuts are applied \cite{Binoth:2005ua,Binoth:2006mf}.

In this article we consider the hadronic production of $Z$-boson pairs.
It has been studied extensively in the 
literature including higher order 
corrections \cite{kauer_nikolas:ppZZNLO,Campbell:1999ah}. 
Production of $Z$ boson pairs through gluon fusion 
contributes at ${\cal O}(\alpha_s^2)$ relative to $q\bar{q}$ annihilation, 
but its importance is likewise enhanced by the large gluon flux at the LHC.
It was analyzed in Refs.~\cite{Dicus:1987dj,Glover:1988rg}. 
Leptonic $Z$ decays were subsequently studied for on-shell \cite{Matsuura:1991pj} and 
off-shell \cite{Zecher:1994kb} vector bosons.

Here, we present the first complete calculation of the gluon-induced loop process 
$gg \to Z^\ast(\gamma^\ast)Z^\ast(\gamma^\ast) \to \ell\bar{\ell}\ell'\bar{\ell'}$,
allowing for arbitrary invariant masses of the $Z$ bosons and including the 
$\gamma$ contributions.
Our calculation employs the same methods as Refs.~\cite{Binoth:2005ua,Binoth:2006mf}.
The tensor reduction scheme of Refs.~\cite{Binoth:1999sp,Binoth:2005ff} has been 
applied to obtain one amplitude representation implemented in our program.  
We compared it numerically with an amplitude representation based on FeynArts/FormCalc 
\cite{Hahn:1998yk,Hahn:2000kx} and found agreement.
Note that single resonant diagrams (in the case of massless leptons) and 
the corresponding photon exchange diagrams give a vanishing contribution.
A combination of the multi-channel \cite{Berends:1994pv} and phase-space-decomposition \cite{kauer_nikolas:psdecomp} Monte Carlo
integration techniques was used with appropriate mappings to
compensate peaks in the amplitude.


\section{Parton-level results}

In Ref.~\cite{Dawson:2008uu} we presented numerical results for the process $pp \to
Z^\ast(\gamma^\ast)Z^\ast(\gamma^\ast) \to \ell\bar{\ell}\ell'\bar{\ell'}$ at the LHC,
i.e.~for the production of two charged lepton pairs with different flavor\footnote{%
Note that no flavor summation is applied.}
focusing on resonant $Z$-pair production and decay by applying the window cut $75$ GeV $< M_{\ell\ell} < 105$ GeV 
to the invariant masses of $\ell\bar{\ell}$ and $\ell'\bar{\ell'}$, which suppresses 
the photon contribution to less than 1\%.  
One finds that enhanced by the large gluon flux at the LHC
the $gg$ process yields a 14\% correction to the total $ZZ$ cross section calculated 
from quark scattering at NLO QCD.  Relative to the LO $q\bar{q}\to ZZ$ 
prediction the $gg$ contribution is about $20\%$ (in agreement with 
Ref.~\cite{Zecher:1994kb}).
The remaining theoretical uncertainty introduced by the QCD scale was estimated 
by varying the renormalization and factorization scales independently 
between $M_Z/2$ and $2M_Z$.  For the gluon fusion process we found a renormalization 
and factorization scale uncertainty of approximately $20\%$. The scale
uncertainty of the $q\bar{q}\to ZZ$ process at NLO is approximately
$4\%$.  
In addition to cross sections for the LO, NLO QCD and $gg$ processes, the 
distributions in the invariant mass $M_{4l}$ of the four produced leptons and 
the pseudorapidity of the negatively charged lepton are also shown in 
Ref.~\cite{Dawson:2008uu}.

For Higgs masses below the $Z$-pair threshold, the virtual photon contribution to the 
$Z^\ast(\gamma^\ast)Z^\ast(\gamma^\ast)$ background cannot be neglected, since 
almost always one of the produced $Z$ bosons will be off resonance.
We thus present numerical results calculated with a minimal set of cuts,
i.e.~only $M_{\ell\bar{\ell}}, M_{\ell'\bar{\ell'}} > 5$ GeV in order to exclude the photon 
singularity, and using the following set of input
parameters:
$M_W = 80.419$ GeV,  
$M_Z = 91.188$ GeV, 
$G_\mu  = 1.16639 \times 10^{-5}$ GeV$^{-2}$,
$\Gamma_Z  = 2.446$ GeV.
The weak mixing angle is given by $c_{\rm w} = M_W/M_Z,\ s_{\rm w}^2 =
1 - c_{\rm w}^2$.  The electromagnetic coupling is defined in the
$G_\mu$ scheme as $\alpha_{G_\mu} = \sqrt{2}G_\mu M_W^2s_{\rm
  w}^2/\pi$.  The masses of external fermions are neglected. The
values of the heavy quark masses in the intermediate loop are set to
$M_t = 170.9$~GeV and $M_b = 4.7$~GeV.  The $pp$ cross
sections are calculated at $\sqrt{s} = 14$~TeV employing the CTEQ6L1
and CTEQ6M \cite{Pumplin:2002vw} parton distribution functions at
tree- and loop-level, corresponding to $\Lambda^{\rm LO}_5 = 165$ MeV
and $\Lambda^{\overline{{\rm MS}}}_5 = 226$ MeV with one- and two-loop
running for $\alpha_s(\mu)$, respectively.  The renormalization and
factorization scales are set to $M_Z$.

\begin{figure}[htb]
\vspace*{0.2cm}
\centerline{\includegraphics[height=7.8cm,angle=90,clip=true]{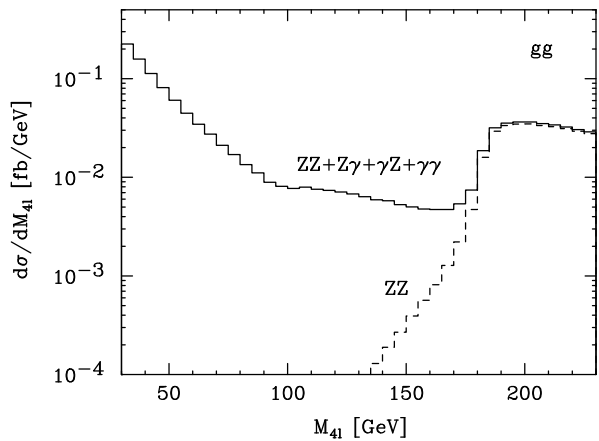}}
\caption{%
  Distribution in  the $\ell\bar{\ell}\ell'\bar{\ell'}$ invariant
  mass $M_{4l}$ for the gluon scattering process 
  $gg \to Z^\ast(\gamma^\ast)Z^\ast(\gamma^\ast) \to \ell\bar{\ell}\ell'\bar{\ell'}$ 
  at the LHC with $Z^\ast Z^\ast$ contribution only (dashed) and all
  contributions (solid).
  Other details as in Table \protect\ref{kauer_nikolas_tbl:xsec}.
  \label{kauer_nikolas_fig:basiccuts}}
\end{figure}

In Table~\ref{kauer_nikolas_tbl:xsec} we compare cross sections 
for $\ell\bar{\ell}\ell'\bar{\ell'}$ production in gluon 
scattering with LO and NLO results for the quark scattering processes at the LHC.\footnote{Since 
we are interested in $Z^\ast(\gamma^\ast)Z^\ast(\gamma^\ast)$ production as a background, 
the $gg\to H \to ZZ$ signal amplitude is not included.} The LO and NLO quark scattering
processes are computed with MCFM \cite{Campbell:1999ah}, which
implements helicity amplitudes with full spin correlations
\cite{Dixon:1998py} and includes finite-width effects and
single-resonant corrections.  The gluon fusion process is calculated with our 
program \texttt{GG2ZZ} \cite{Dawson:2008uu,GG2ZZ}.
\begin{table}[htb]
\centerline{
\def\arraystretch{1.5}
\begin{tabular}{|c|cc|c|c|}
\cline{1-3}
\multicolumn{3}{|c|}{$\sigma(pp \to Z^\ast(\gamma^\ast)Z^\ast(\gamma^\ast) \to \ell\bar{\ell}\ell'\bar{\ell'})$~[fb]} & \multicolumn{1}{c}{} & \multicolumn{1}{c}{}\\ \hline
& \multicolumn{2}{|c|}{\raisebox{1ex}[-1ex]{$q\bar{q}$}}
& \multicolumn{1}{c|}{} &\multicolumn{1}{c|}{} \\[-1.5ex]
\cline{2-3}
\multicolumn{1}{|c|}{\raisebox{2.7ex}[-2ex]{$gg$}} & 
\raisebox{0.9ex}{LO} & \raisebox{0.9ex}{NLO} 
& \raisebox{2.25ex}[-2ex]{$\frac{\sigma_{\rm NLO}}{\sigma_{\rm LO}}$} & 
  \raisebox{2.25ex}[-2ex]{$\frac{
 \sigma_{{\rm NLO}+gg}}{\sigma_{\rm NLO}}$}
\\[-1.5ex]
  \hline
 $16.3(1)$ & $105.2(1)$ & $118.9(2)$ & 1.13 & 1.14 \\
 \hline
\end{tabular}}
\vspace*{.5cm}
\caption{\label{kauer_nikolas_tbl:xsec}
  Cross sections for the gluon and quark scattering contributions to
  $pp \to Z^\ast(\gamma^\ast)Z^\ast(\gamma^\ast) \to \ell\bar{\ell}\ell'\bar{\ell'}$ at the LHC
  ($\sqrt{s} = 14$ TeV), where a minimal cut $M_{\ell\bar{\ell}}, M_{\ell'\bar{\ell'}} > 5$ GeV is applied. 
  The integration error is given in brackets. We also show the ratio of the NLO to LO 
  cross sections and the ratio of the combined NLO+$gg$ contribution to the NLO cross 
  section.  Input parameters are defined in the main text.}
\end{table}
For $pp\to Z^\ast(\gamma^\ast)Z^\ast(\gamma^\ast)\to \ell\bar{\ell}\ell'\bar{\ell'}$ 
we find a NLO $K$-factor of 1.13 when only a $M_{\ell\bar{\ell}}, M_{\ell'\bar{\ell'}} > 5$ GeV cut is applied.
The $gg$ process yields an additional correction of $14\%$ relative to the NLO 
prediction for the $q\bar{q}$ process.
In Fig.~\ref{kauer_nikolas_fig:basiccuts}, invariant mass $M_{4l}$ 
distributions for the $gg$ subprocess are compared by taking into account
only the $Z^\ast Z^\ast$ contribution as well as all contributions.
We observe that for Higgs masses below the $Z$-pair threshold, where 
one $Z$ boson is produced off-shell, the photon contribution to the background 
is important.


\section{Conclusions}
We have calculated the loop-induced gluon-fusion process 
$gg \to Z^\ast(\gamma^\ast)Z^\ast(\gamma^\ast) \to \ell\bar{\ell}\ell'\bar{\ell'}$, 
which provides an important background for Higgs boson searches in the $H \to ZZ$
channel at the LHC. 
Our calculation demonstrates that the photon contribution is important for Higgs 
masses below the $Z$-pair threshold.  The $gg$-induced process yields a 
correction of about $15\%$ relative to the NLO QCD prediction for the 
$q\bar{q}$-induced process when only a $M_{\ell\bar{\ell}}, M_{\ell'\bar{\ell'}} > 5$ GeV cut is 
applied.
We conclude that the complete gluon-gluon induced background process
should be taken into account for an accurate determination of the 
discovery potential of Higgs boson searches in the $pp\to H\to ZZ \to$ 
leptons channel if $M_H<2M_Z$.

\section*{Acknowledgements}

This work was supported by the BMBF and DFG, Germany (contracts 05HT1WWA2
and BI 1050/2).
T.B.~and N.K.~thank the Galileo Galilei Institute for Theoretical
Physics for the hospitality and the INFN for partial support during the
completion of this work.

\begin{footnotesize}

\end{footnotesize}

\end{document}